\documentclass[a4paper,11pt]{article}
\pdfoutput=1 
\usepackage{jinstpub} 
\usepackage{subcaption}
\usepackage{lineno}

\title{Medipix3 for dosimetry and real-time beam monitoring: first tests at a 60 MeV proton therapy facility}
\author[a,b,c,1]{J. S. L. Yap \note{Corresponding author.}}
\author[d]{N. J. S. Bal}
\author[e]{A. Kacperek}
\author[b,c,f]{J. Resta L\'opez}
\author[b,c]{C. P. Welsch}

\affiliation[a]{The University of Melbourne, Melbourne, Australia.}
\affiliation[b]{The University of Liverpool, Liverpool, UK}
\affiliation[c]{Cockcroft Institute, Warrington, UK}
\affiliation[d]{Nikhef, Amsterdam, Netherlands}
\affiliation[e]{Clatterbridge Cancer Centre, NHS Foundation Trust, Wirral, UK}
\affiliation[f]{Institute of Materials Science (ICMUV), University of Valencia, Valencia, Spain}

\emailAdd{jacinta.yap@unimelb.edu.au}

\abstract{Charged particle therapy (CPT) is an advanced modality of radiation therapy which has grown rapidly worldwide, driven by recent developments in technology and methods of delivery. To ensure safe and high quality treatments, various instruments are used for a range of different measurements such as for quality assurance, monitoring and dosimetry purposes. With the emergence of new and enhanced delivery techniques, systems with improved capabilities are needed to exceed existing performance limitations of conventional tools. The Medipix3 is a hybrid pixel detector able to count individual protons with millisecond time resolution at clinical flux with near instant readout and count rate linearity. The system has previously demonstrated use in medical and other applications, showing wide versatility and potential for particle therapy. In this work we present measurements of the Medipix3 detector in the 60 MeV ocular proton therapy beamline at the Clatterbridge Cancer Centre, UK. The beam current and lateral beam profiles were evaluated at multiple positions in the treatment line and compared with EBT3 Gafchromic film. The recorded count rate linearity and temporal analysis of the beam structure was measured with Medipix3 across the full range of available beam intensities, up to $3.12 \times 10^{10}$ protons/s. We explore the capacity of Medipix3 to provide non-reference measurements and its applicability as a tool for dosimetry and beam monitoring for CPT. This is the first known time the performance of the Medipix3 detector technology has been tested within a clinical, high proton flux environment.}


\begin{document}
\maketitle 
\flushbottom

\section{Introduction}
\label{Introduction}
The use of particle beams for radiotherapy is expanding worldwide, led by developments in related technologies, growing clinical practice and reported improvements in treatment outcomes. The physical advantages offered by ions as demonstrated by the characteristic `Bragg Peak', translate to a higher possibility of delivering a more precise amount of radiation with greater radiobiological effectiveness. The exploitation of these benefits have progressed significantly in recent years: advanced delivery techniques and modalities such as proton beam therapy (PBT) are well established. 

To ensure the safe and effective deliver of charged particle therapy, it is essential that the characteristics of the beam are accurately and reliably measured. Various systems are used for different measurements for beam monitoring, quality assurance as well as dosimetry under reference or non-reference conditions. Procedures may vary across facilities as they depend on accelerator type, delivery method, vendor, regulations and other parameters. Dosimetry practices are well defined as the determination of dose must be precise and reproducible. The recommended protocol for absolute, relative, reference and non-reference measurements adopted for CPT are detailed in reports \cite{TRS398,DeLuca2007}; the general standards for existing detectors are discussed extensively in literature and summarised in \cite{Seco2014, Giordanengo2017, Karger2010, Giordanengo2018, AAPM-TG224}. Each detector has certain advantages and disadvantages given the measurement application and no single tool can characterise all the necessary beam parameters: multiple detectors for different measurements are required in the workflow. For dosimetry and beam monitoring, the fundamental technology of conventional tools have remained relatively unchanged for the past 10--20 years \cite{Giordanengo2018}. In contrast, limitations are becoming more pronounced as new techniques are emerging, particularly with the evolution of accelerators and the shift to higher `FLASH' dose rates \cite{Favaudon2014, Jolly2020}. These performance requirements surpass current instruments \cite{Nesteruk2021a}, driving research and development into detector systems with improved capabilities. 

Silicon based detectors have been utilised mainly due to their high spatial resolution and sensitivity but the typically used diodes and commercially available systems exhibit issues with energy dependence and a poor response at higher dose rates, along with other disadvantages \cite{Grusell2000}. Nevertheless, there are many recognised beneficial properties and solid state devices have been widely investigated, resulting in developments into several novel instruments \cite{Bruzzi2016, Rosenfeld2016, Schnuerer2018a, Vignati2017, Taylor2016} for a range of applications in radiotherapy. Similarly, the use of hybrid pixel detectors has demonstrated great versatility and has been successfully applied for a broad range of areas such as astrophysics, dosimetry, electron microscopy, life sciences, etc. The Medipix detector technology was originally developed for particle tracking at the LHC and then used for radiation imaging and x-ray detection \cite{Ballabriga2020}. The newer generations have further progressed its specific use for medical applications and suitability for CPT \cite{Bisogni2009, Rosenfeld2020}.

The Medipix3 is a hybrid pixel detector which comprises a single quantum counting chip and a SPIDR (Speedy PIxel Detector Readout) readout system, enabling direct measurements by individual particle counting with relatively high count rate and good radiation hardness \cite{BALLABRIGA2011S15}. The chip is capable of high flux operation, thus making it appropriate for the characterisation and measurement of ion beams. In this paper we present test results of the system within a clinical PBT environment for the first time, to examine the potential and applicability of the Medipix3 detector as a tool for dosimetry and monitoring for CPT beams.


\section{Materials and Methods}
Measurements were performed with the 60 MeV clinical proton therapy beam at the Clatterbridge Cancer Centre (CCC), UK. The Medipix3 system was placed at three different locations in the treatment beamline and irradiated simultaneously with GAFchromic\textsuperscript{TM} EBT3 film \cite{EBT3} using beam currents ranging from 0.012--1.97 nA. The transverse beam profiles obtained with film were determined and compared with the hit distribution recorded by the detector. The average count rate for each run was also examined to determine the count rate linearity of the response of the detector. 

\subsection{Clatterbridge proton beam}
The Clatterbridge facility is the world’s first hospital based particle therapy centre and has been successfully treating patients for ocular cancers since 1989 \cite{Kacperek2009}. The accelerator and beamline was originally built and commissioned for fast neutron therapy trials but was converted shortly after into a proton therapy service, specifically for the treatment of eye tumours. Given its pioneering and historical operation, many of the treatment line components and quality assurance (QA) equipment were constructed in-house. Several modifications to the accelerator and transport line were necessary and performed over time, resulting in differences to characteristics and properties of the beam reported initially at commissioning. The current state of the facility has been studied in detail in \cite{Yap2020}, general characteristics and treatment beam parameters are mentioned; for this study, dose rates below and within the clinical range were used. 

As the facility operates a busy patient schedule, the delivery system is optimised and designed in such a way to consistently deliver a beam constrained by clinical requirements. Basic beam measurements and dose monitoring are performed within the treatment beamline itself: the integrated charge is provided by the electrometer attached to the second scattering foil, a pair of ionisation chamber dose monitors (F) give the converted dose in monitor units and a set of tungsten cross-wires (G) are used for patient positioning (imaged with x-ray panels). An X-Y diode scanner (not shown), placed after the treatment nozzle, is used to check the dose uniformity in both directions in the transverse plane before each treatment fraction. As shown below in figure \ref{fig:deliverysys}, a resulting conformal dose distribution is delivered by several beam shaping elements in the passive scattering system: (A) two tungsten foils and a brass stopper moderate the shape and fluence of the beam, (B $\&$ C) range shifters vary the depth, (E) multiple beamline anti-scatter collimators restrict the transverse spread and a patient specific collimator constructed to match the tumour shape in the axial direction, can be placed within the (H) brass treatment nozzle.

\begin{figure}[htb]
    \centering
    \includegraphics*[width=\textwidth]{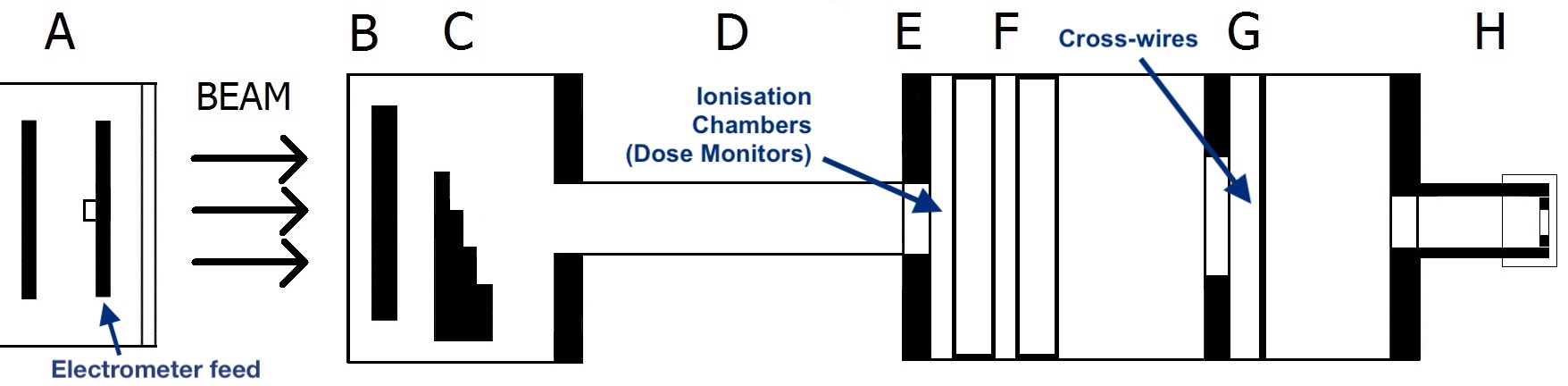}
    \caption{Schematic of the CCC treatment beamline with diagnostic components (electrometer, dose monitors and cross-wires) and delivery elements (A-H; scattering foils, modulators, nozzle etc.) labelled.}
    \label{fig:deliverysys}
\end{figure}

The CCC treatment line provides unique clinical conditions which are not common by modern standards; the passive delivery system as well as resulting uncertainties associated with the beam parameters and quality presented both a challenging and promising environment for testing with the Medipix3 system. In addition, these measurements also provide further information about the behaviour and parameters of the Clatterbridge beam. It is noted that there are ongoing experiments and simulation studies being performed on the beamline \cite{YapTopas,UCLTOPAS,Vitti2020}. Information on the transverse beam profiles, beam divergence and lateral spread are also useful for model verification and validation. 


\subsection{Medipix3 detector}
The Medipix3 is a versatile and fast hybrid pixel application specific integrated circuit (ASIC). Among its features are configurable pixel pitch between 55 \textmu m and 110 \textmu m, relatively high count rates of up to $\sim$100 kHz per pixel ($\sim$6.5 GHz over one chip if evenly distributed) and it can use many different sensor types of various thicknesses. Common sensor options include high-resistivity silicon (Si), gallium arsenide (GaAs), cadmium telluride (CdTe), CVD (chemical vapor deposition) diamond and even certain gases.

\begin{figure}[htb]
    \centering
    \includegraphics*[width=0.97\textwidth]{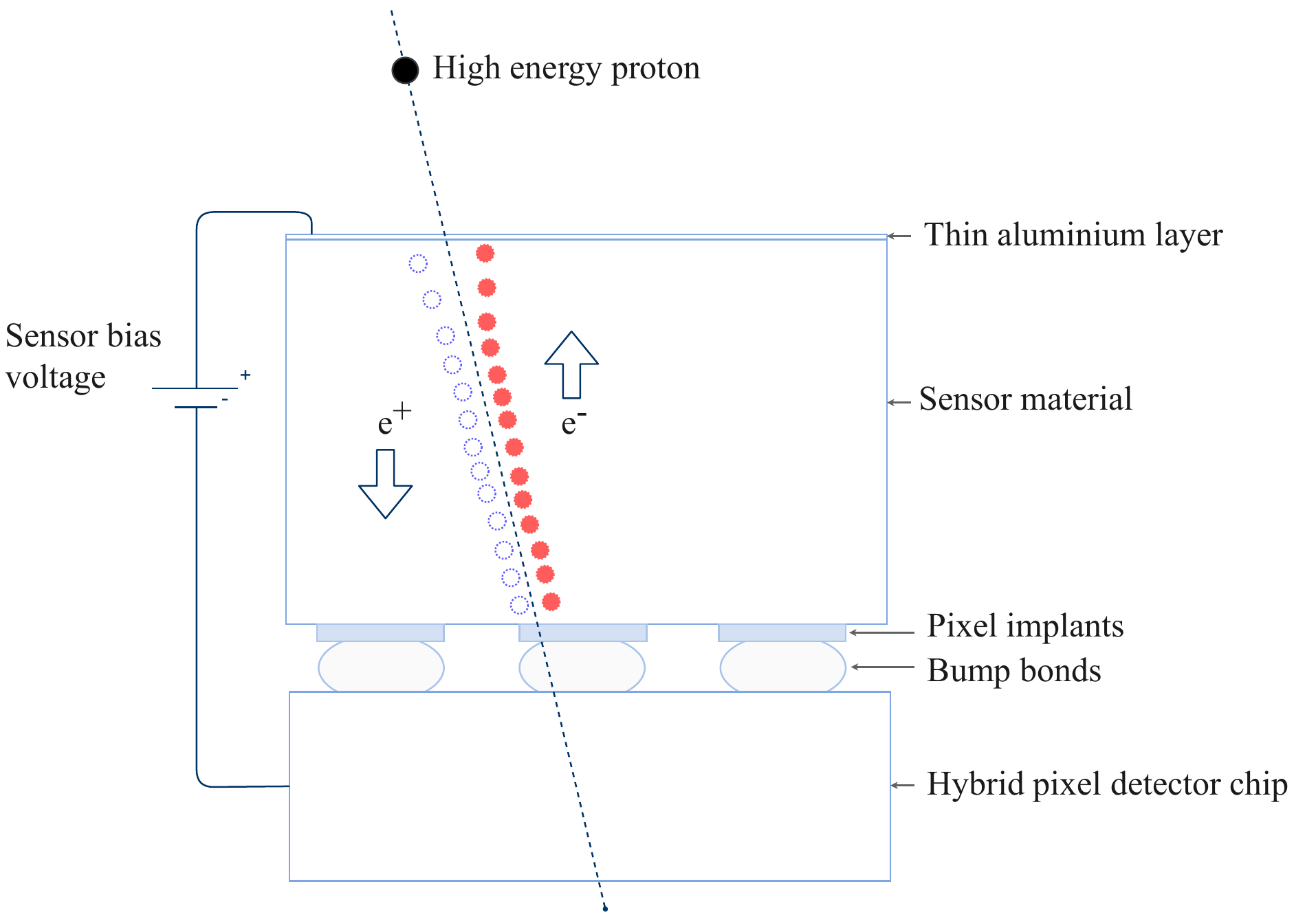}
    \caption{ An overview of a hybrid pixel detector when a high energy proton enters the sensor bulk and transfers energy mainly by ionising sensor atoms producing electron-hole pairs which drift to the cathode and anode depending on the polarity of the induced electric field. }
    \label{fig:hybrid-pixel-detector-proton}
\end{figure}

As illustrated in figure \ref{fig:hybrid-pixel-detector-proton}, when a 60 MeV proton impinges on the sensor and travels through the sensor bulk, it produces many electron-hole pairs by electronic and nuclear interactions where the mean rate of energy loss is described by the Bethe-Bloch equation. The generated electron-hole pairs drift to the front and back-side of the sensor due to an applied electric field. The moving charge induces a current (Shockley-Ramo theorem) through the pixel implants (collection electrodes) which are reverse biased, depleted p-n junctions. This starts the pulse processing chain within one pixel. The measured charge is proportional to the energy deposition of the particle. Since the mean range of a 60 MeV proton in Si is $\frac{\textrm{CSDA range}} {\rho_{Si}} = \frac{3.94~g/cm^{2}}{2.33~g/cm^{3}} = 16.93$ mm and the detector is 500 \textmu m thick, the Bragg peak does not occur within the silicon bulk of the detector if the detector is perpendicular to the beam.\footnote{CSDA proton range in Si is from the PSTAR database, NIST \cite{PSTARNIST}}. If the detector is positioned parallel to the beam, the Bragg peak for 60 MeV protons would then be within the width of the detector; the feasibility to perform depth-dose profile measurements also presents a further avenue to explore in future.

Nevertheless, at this proton energy, it is likely that the deposited charge will be collected by more than one pixel. This effect is called `charge sharing' \cite{Gimenez_2011}. It results in the detector counting an average of more than once per 60 MeV proton. A detector mode exists to address charge sharing called the `charge summing mode' which sums the collected charge over an arbitrary 2 $\times$ 2 pixel grid. The penalty is a reduction of approximately one order of magnitude in count rate along with double the electronic noise.

In addition to simple geometric effects, it is likely that the charge cloud generated by the protons will also not be centred on the pixels and therefore cause an increased count in comparison to the number of protons which actually pass through the sensor. Charge summing mode was not used in this work due to the significant count rate penalty.

A detailed description of the Medipix3 chip can be found in \cite{BALLABRIGA2011S15}. For this study, the ASIC was bump-bonded to a 500 \textmu m silicon sensor with a pixel pitch of 55 \textmu m. This detector is made of four 55 $\times$ 55 \textmu m$^{2}$ chips in a 2 $\times$ 2 arrangement, where each chip consists of 256 $\times$ 256 = 65,536 pixels. Therefore with 4 chips, the total number of square pixels is 262,144 and the active area covers 28 $\times$ 28 mm$^{2}$.

\begin{figure}[htb]
    \centering
    \includegraphics*[width=\textwidth]{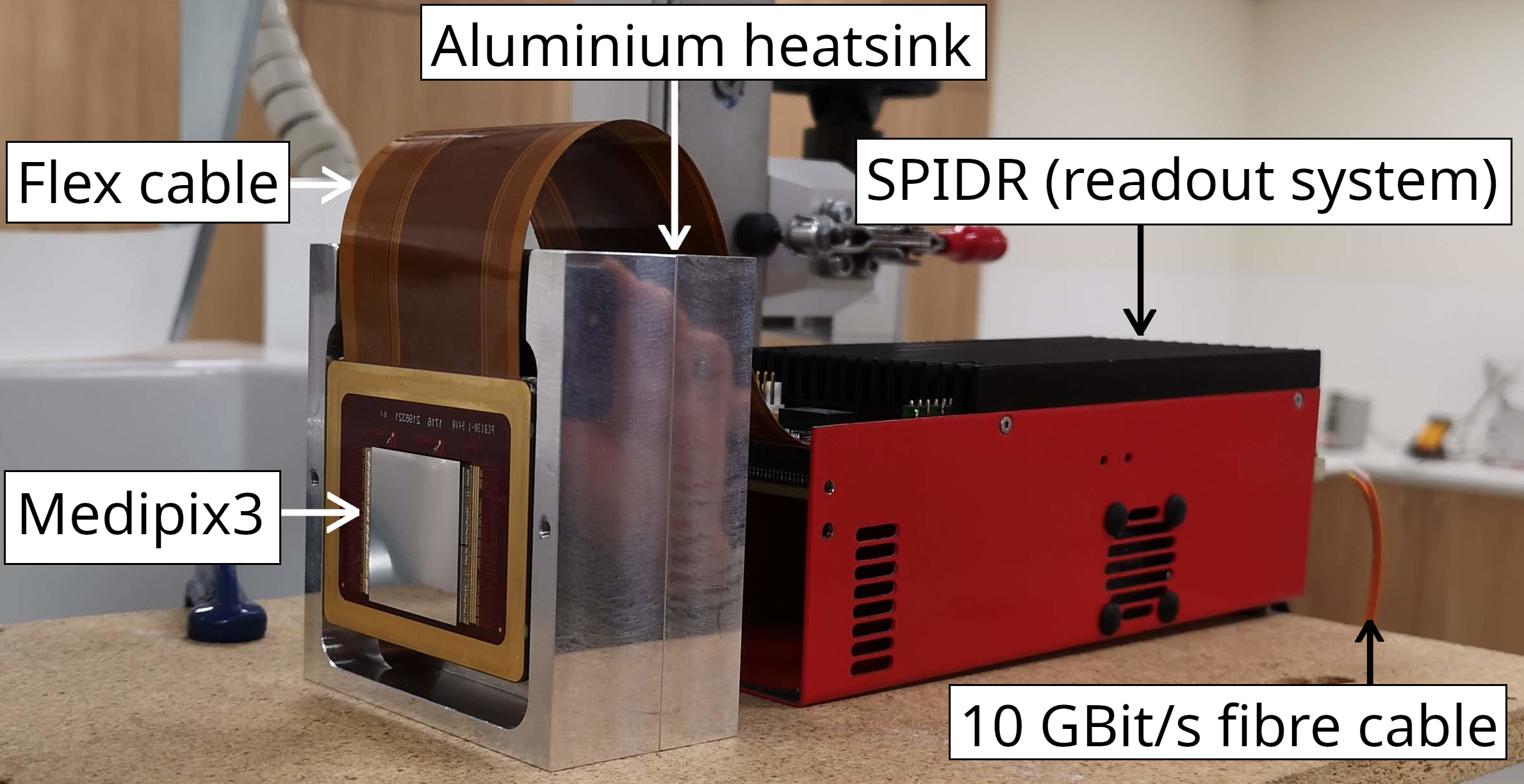}
    \caption{Medipix3 chipboard affixed to an aluminium cooling block with the red SPIDR readout system behind. The 10 GBit/s fibre cable takes the data from the detector to the readout computer and a power cable is also attached to the SPIDR (not in view).}
    \label{fig:labelledMP3}
\end{figure}

These measurements were performed with a Medipix3 based detector using a SPIDR readout system \cite{Visser_2015, Heijden_2017} from Nikhef, with the experimental layout as shown in figure~\ref{fig:setup}. The detector was biased with +100 V and configured to collect holes. The sensitive area of the detector was held in an aluminium cooling block and connected to the readout system by fibre cables. A small assembly was designed to securely hold a piece of film 3.5 cm in front of and parallel to the sensor and was clamped onto the edges of the block (figure \ref{fig:labelledMP3}).
 
\subsection{Experimental setup}
\begin{figure}[htb]
    \centering
    \includegraphics*[width=\textwidth]{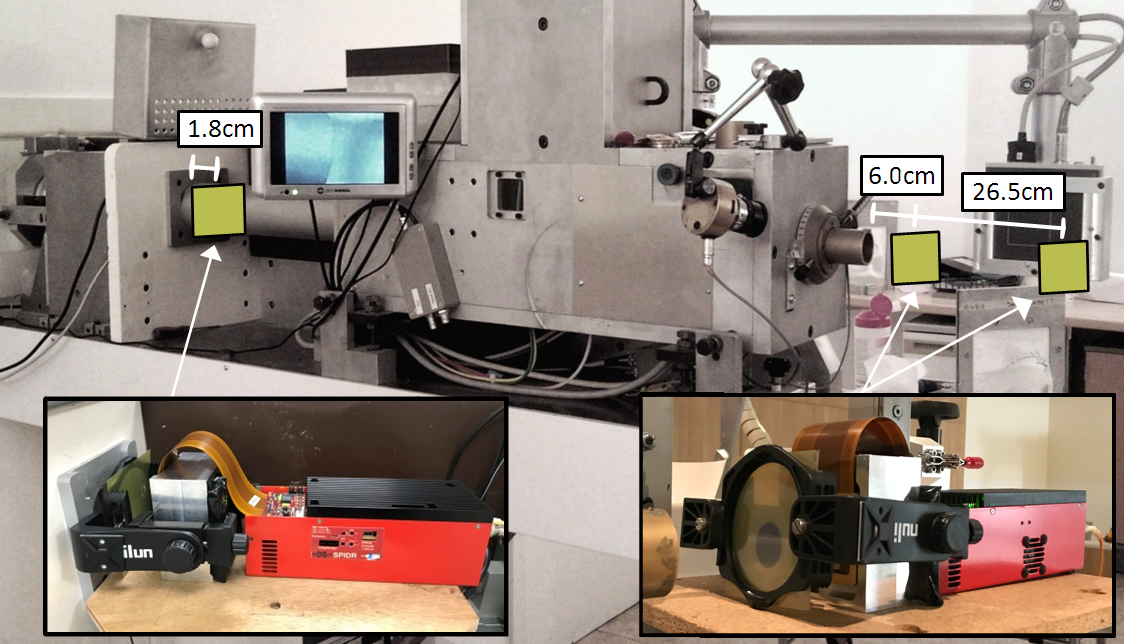}
    \caption{Experimental setup with measurements performed at three different locations (1.8 cm in the integration zone, 6 cm and 26.5 cm after the treatment nozzle) with the Medipix3 detector and EBT3 film simultaneously. A clamp assembly holds the film 3.5 cm directly upstream of the detector face.}
    \label{fig:setup}
\end{figure}

The Medipix3 detector was placed at three locations throughout the treatment beamline and irradiated under varying beam conditions (table~\ref{tab:run-data}). In the integration zone, downstream of the scattering foils and modulation box (aluminium drift pipe removed) and two positions downstream of the treatment nozzle. Sections of EBT3 Gafchromic film were also positioned in front of the detector (figure~\ref{fig:setup}) and irradiated simultaneously for direct comparison. 

\begin{table}[htb]
\centering
\caption{Overview measurement parameters of detector position, acquisition time and beam currents.}
\begin{tabular}{|c|c|c|c|}
\hline
Run (\#) & Beam current (nA) & Time (s) & Distance from nozzle (cm) \\ \hline
2        & 0.012             & 97.2     & 9.5                       \\ 
3        & 0.052             & 99.8     & 9.5                       \\ 
4        & 0.35              & 49.8     & 9.5                       \\ 
5        & 0.69              & 44.9     & 9.5                       \\ 
6        & 0.27              & 32.6     & 9.5                       \\ 
7        & 0.27              & 29.5     & 30.0                      \\
8        & 1.35              & 75.0     & 30.0                      \\
9        & 1.35              & 68.9     & 9.5                       \\ 
10       & 1.97              & 66.4     & 9.5                       \\ 
11       & 0                 & 9.0      & 9.5                       \\
14       & 2.2               & 103.0    & Integration zone          \\ 
15       & 2.1               & 285.3    & Integration zone          \\ \hline
\end{tabular}
\label{tab:run-data}
\end{table}

Prior to the experiment, the estimated dose and fluence were calculated to determine the range of measurements and expected number of registered events. The Medipix3 chip can tolerate relatively high amounts of radiation (still functional after irradiation by 1 MeV neutron equivalent fluence of 5 $\times 10^{14}$ cm$^{-2}$ \cite{Akiba2016}) and hence was also situated in the beginning of the integration zone. At this position, the beam FWHM was determined to be smaller than the sensor sensitive area however diverges significantly along the  beamline. A 20 mm collimator was placed within the treatment nozzle for the downstream measurements, to ensure complete beam coverage across the sensor.

A minimum stable beam current of 0.012 nA was attained on the day and then ramped up to higher currents for subsequent measurements. This is measured by the electrometer connected to the second scattering foil and is listed in table~\ref{tab:run-data} for each run for consistency: the dose monitors could not provide readings whilst the detector was placed within the integration zone. The foil currents have a linear relationship with the ion chamber monitor units. 


\subsection{EBT3 film}
\begin{figure}[htb]
    \centering
    \includegraphics*[width=0.7\textwidth]{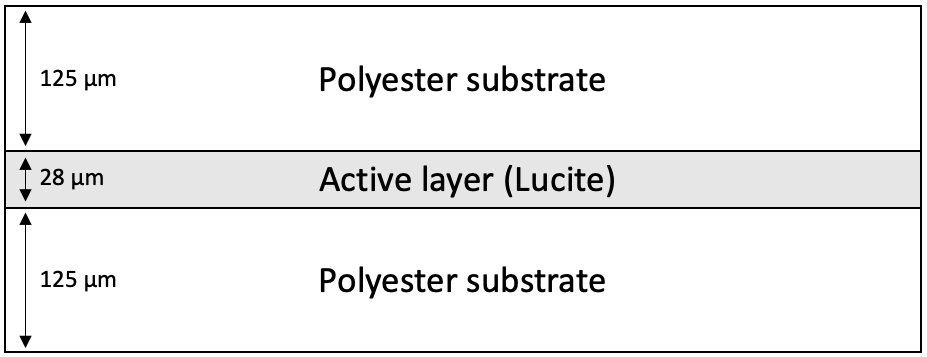}
    \caption{Sketch of the composite layers in EBT3 film, a 28 \textmu m active region is surrounded by two thicker substrate layers.}
    \label{fig:EBT3}
\end{figure}

EBT3 GAFchromic film is a standard radiochromic film dosimeter commonly used for quality assurance in radiation therapy to analyse the geometrical beam characteristics (i.e. uniformity, shape and size) and measure the 2D or 3D dose distribution \cite{Castriconi2017,Troja2000}. For patient specific or machine verification, it is essential to be able to perform checks with high accuracy and reliability, prior to delivering a course of treatment. In proton beam therapy, the use of Gafchromic EBT3 for film dosimetry to determine beam performance and quality is well established \cite{Reinhardt2012, Sorriaux2013, Devic2016}. EBT3 Gafchromic film is made of a 28 \textmu m layer of Lucite, enclosed by 125 \textmu m  of polyester substrate on each side (figure~\ref{fig:EBT3}). 

\begin{figure}[htb]
    \centering
    \includegraphics*[width=\textwidth]{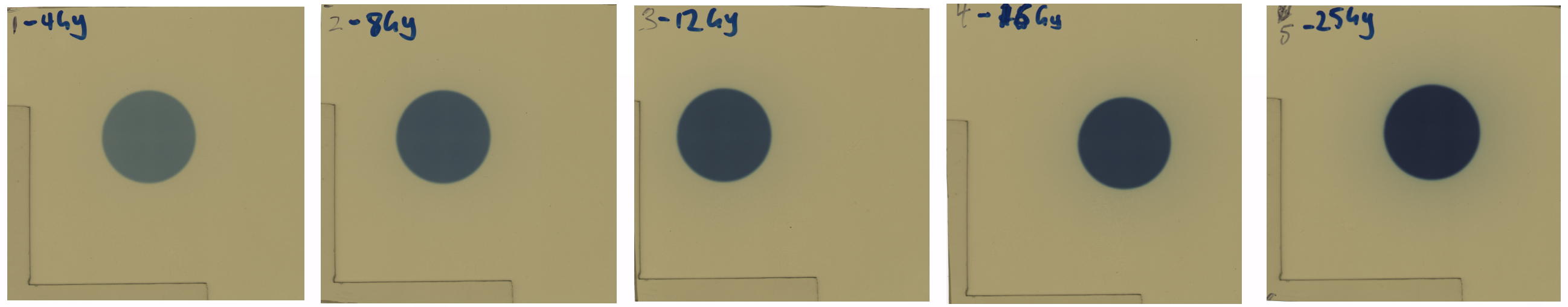}
    \caption{Calibration set of film irradiation with 4--25 Gy.}
    \label{fig:FilmAll}
\end{figure}

Exposure to ionising radiation results in polymerisation of free radicals within the active layer, inducing the film to darken \cite{Battaglia2016}. EBT3 self develops and the dark colouring or optical density (OD) is proportional to the extent of irradiation, increasing with absorbed dose. The geometrical beam distribution is provided as a function of the dose, evaluated by converting the grey value of each pixel to an OD value. The high spatial resolution and low energy dependence enables measurements of the transverse dose profiles. However, the use of EBT3 film in CPT is limited due to significant quenching effects and saturation at points of high linear energy transfer, such as at the Bragg peak \cite{Castriconi2017, Yonai2018}. These effects were not experienced as these were transmission measurements. It is also noted that there are multiple considerations and associated sources of uncertainties \cite{LeonMarroquin2016}. Calibration measurements must be performed under specific conditions: there are established methods and standard protocol for the OD to dose conversion process as widely reported in literature \cite{Castriconi2017, Vadrucci2015, Devic2004, Sorriaux2013}. These measurements were carried out in accordance with these methods; the OD value for each pixel results in a corresponding dose (Gy) and is determined by establishing a correlation with known quantities of radiation. This is done by exposing a calibration set of film (figure~\ref{fig:FilmAll}) to well defined quantities of radiation under standard conditions, with reference to the dose measured by the ion chambers. 

The calibration set of film was placed downstream of the nozzle and irradiated individually with doses ranging from 4--25 Gy to generate a calibration curve. For the film attached to the detector system (figure~\ref{fig:setup}), a single piece was cut into several equivalent segments and labelled to ensure that the direction and orientation remained consistent. Following complete development ($>$24 hours), the irradiated film pieces were scanned using an EPSON 750 scanner and saved as 48-bit TIFF (Tag Image File Format) images with no colour corrections at 150 dots per inch (DPI). All film analysis was done using the image processing software ImageJ \cite{ImageJ}. A circular region of interest (ROI) was selected such that it was encased inside each beam spot and duplicated for each film. ImageJ was used to generate the ROI intensity metrics for each colour channel to obtain a calibration curve, providing a correlation between the OD and dose.

\subsection{Film calibration and image analysis}
\label{FilmAnalysis}
The calibration curve is obtained by evaluating the net OD values \cite{Devic2004} across the full dose range, where the film response to the measured dose is expressed as the difference between transmission intensities:  

\begin{equation}
net \: OD = OD_{exp} - OD_{unexp} = \log _{10}\left(\frac{I_{unexp}-I_{bckg}}{I_{exp }-I_{bckg}}\right),
\label{E_netOD}
\end{equation}

\noindent where \textit{exp} refers to whether the film was irradiated (\textit{unexp}, unexposed) and \textit{bckg} is the zero-light transmission quantity. This is the pixel value related to the white light value of the scanner used. \textit{I} is the respective intensity value and is taken across each colour channel (red, green, blue). The uncertainties and possible errors can be calculated as similarly found in \cite{Devic2004}, where:

\begin{equation}
\sigma_{net \: OD} = \frac{1}{\ln (10)} \sqrt{\left(\frac{\sigma_{unexp}^{2}+\sigma_{ {bckg }}^{2}}{\left(I_{unexp}-I_{bckg}\right)^{2}}+\frac{\sigma_{exp}^{2}+\sigma_{bckg }^{2}}{\left(I_{exp}-I_{bckg}\right)^{2}}\right)}.
\label{E_netODerror}
\end{equation}

\begin{figure}[htb]
    \centering
    \includegraphics*[width=0.78\textwidth]{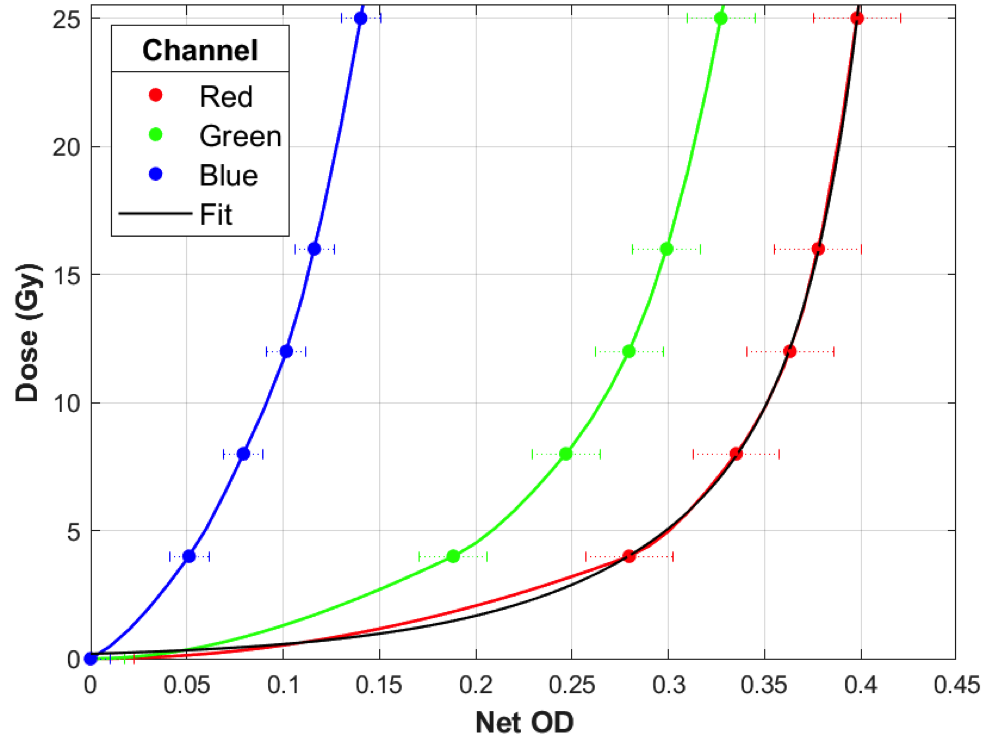}
    \caption{Calibration curve for conversion of net OD (optical density) to dose (Gy).}
    \label{fig:calibrationcurve}
\end{figure}

The net OD values are plotted against the corresponding irradiated doses to determine the response curve as shown in figure~\ref{fig:calibrationcurve}. A calibration curve for each colour channel is obtained and given standard protocol, only the red channel is considered for this case. A curve fit was applied which enables the grey values from the irradiated films to be converted to dose and plotted against position to obtain the transverse beam distributions. Several scripts were developed in Matlab to automate this conversion process which are documented and accessible from \cite{YapFILM}.

The Medipix3 images were obtained by integrating the counts over all frames for each run and also evaluated using ImageJ. Minor artefacts in the central and cross pixels are observed from the interface between the four chips resulting in a larger than average effective pixel size for the cross pixels. Changes for the different grey value range and conversion of pixel size to mm were also made. Minimal post-processing was performed for the Medipix3 data, outlying pixels (noisy and dead) were interpolated from neighbouring pixels, the images summed and then a Gaussian blur of sigma equal to 1 pixel is applied in order to reduce pixel-to-pixel gain variations. The pixel values correspond to the number of times a pixel received charge above a low threshold of 5 keV. Due to charge sharing over multiple pixels, each proton produces more than one count on average. 

For simplicity, we directly converted the counts to dose by scaling the pixel values to the calibrated film values in Gy by radial distance across the transverse plane, for each section irradiated at the same experimental location as the detector face. This preserves the linearity of the grey values as the grey pixel intensities correspond to numbers of hits and also correlate the magnitude of hits to a determined quantity: dose. These were scaled to the dose measured by the film and not directly calculated from counts or the resolved beam current recorded by Medipix3 due to the uncertainties with the electrometer. These were further perturbed by beam instabilities, particularly at the low currents during moments where there was a complete loss of beam. This was presumed to be caused by a dropout in the RF supplied to either one of the `dees' of the cyclotron, resulting in a loss of the accelerating field between the two electrodes and therefore a disruption to the beam. Additionally, this may be related to the deflector which has deteriorated with use or from changes to the ion source when the facility concluded neutron therapy trials and was repurposed as a proton therapy clinic \cite{Yap2020}. Maintenance cleaning of the cyclotron tank can also influence beam operation, resulting in changes to the beam characteristics.

\subsection{Detector activation \& temporal analysis}

Activation is the process whereby incident particles transfer energy to target materials via nuclear interactions which results in a different nuclear energy state to before the interaction. These new nuclear states are typically unstable and go on to transition to a lower energy state by emitting energy in the form of gamma photons or can even fission into multiple fragments.

During irradiation with protons and heavy ions, any detector will become activated to some extent. Depending on the detector technology, flux and radiation type, this can result in radiation damage to the detector. This could be observed as variations in gain or noise baselines or the number of unresponsive pixels. It may be possible for the detector to partially or fully recover from such changes with software compensation, by simply leaving the detector for a long time, annealing the sensor material at a higher temperature or other hardware level interventions. 

The mean count rates for each run and between successive runs were calculated by averaging counts over frames where the beam was not active, to get an indication of the levels of activation within the high flux environment. Furthermore, due to the time resolved nature of the Medipix measurements as opposed to integrated over an entire measurement with film, it was possible to assess the beam current stability across a single measurement. This was further investigated by computing the Fourier transform in order to determine the frequency components of the mean counts over time. If the mean count rate purely varied randomly over time, the frequency component distribution would be flat. However, the distinct narrow peaks demonstrate that the mean counts and therefore the beam current oscillated non-randomly over time.


\section{Measurements and Results}

\subsection{Beam profiles}
\label{BeamProfiles}
\begin{figure}[htb!]
    \centering
    \begin{subfigure}[b]{\textwidth}
    \centering
    \includegraphics[width=0.67\textwidth]{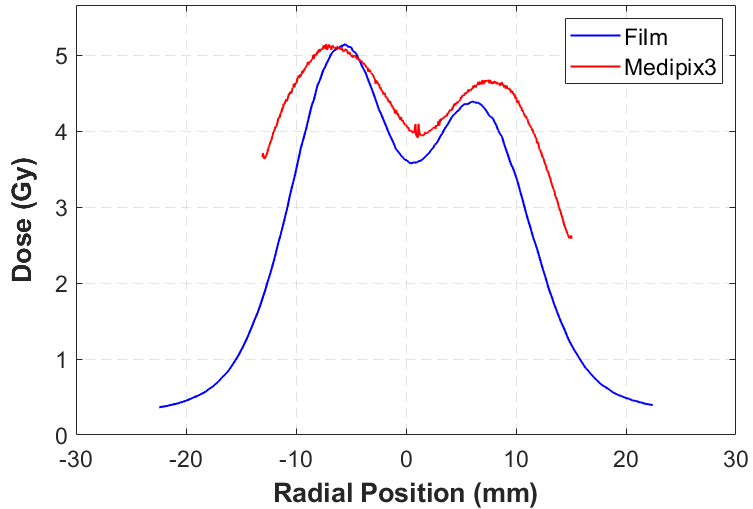}
    \caption{}
    \label{fig:profilesIntegZoneA}
    \end{subfigure}
    \vfil
    \begin{subfigure}[b]{\textwidth}
    \centering
    \includegraphics[width=0.68\textwidth]{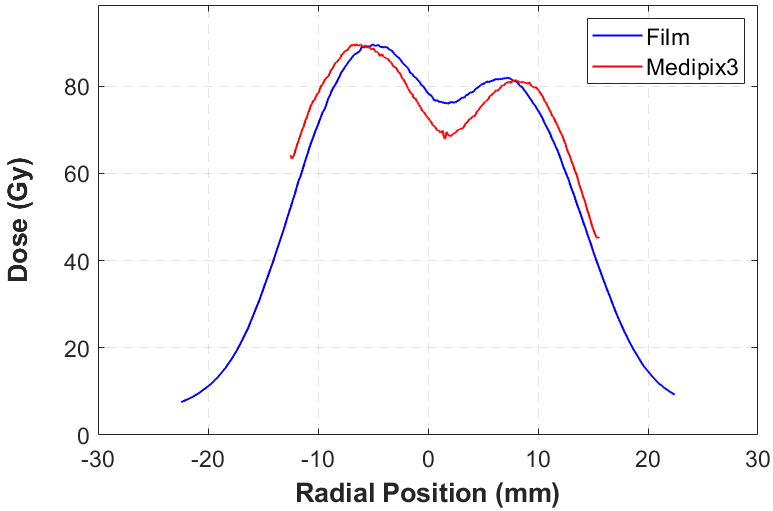}
    \caption{}
    \label{fig:profilesIntegZoneB}
    \end{subfigure}
    \caption{Beam profiles obtained with film and Medipix3 for (a) run 14 and (b) run 15 in the integration zone.}
    \label{fig:profilesIntegZone}
\end{figure}

Beam profiles for each run were obtained and comparisons between both methods are shown for runs 6, 7, 14 and 15. Figure~\ref{fig:profilesIntegZone} shows the transverse distributions measured with the film (blue) and Medipix3 (red) within the integration zone for a standard (run 14, figure~\ref{fig:profilesIntegZoneA}) and at a much higher level (run 15, figure~\ref{fig:profilesIntegZoneB}) of accumulated dose. 

At this location in the beamline, the beam distribution exhibits a double peak. This is caused by the beam stopper attached to the second scattering foil (figure~\ref{fig:deliverysys}) which attenuates a large proportion of the fluence in order to produce a flat, uniform beam at isocentre. For these runs, there is observable correspondence between both methods however the asymmetry of the distributions is distinct. Irradiating the film to a higher accumulated dose (figure~\ref{fig:profilesIntegZoneB}) appears to dilute the film beam distribution toward the directly measured profile by Medipix3. This may be attributed to a smaller range and less variation between the pixel intensity ranges. A corresponding smoothing effect can be seen for the Medipix3 profile. 

The tilt seen in both plots suggests that the whole system was not precisely aligned at the centre of the beam axis but slightly translated. This results in skewed profiles and marked differences between the maximum dose at each apex. For both cases events were detected right through to the edges of the Medipix3 (28 $\times$ 28 mm\textsuperscript{2}) sensor, reflected by the clipped tails along the orthogonal plane. The comparative variations between the two different methods are reasoned to be a consequence of the differences in the position, analysis procedure and fundamental detection processes. In these plots, the Medipix3 measures wider profiles due the 3.5 cm gap between the film and detector. At this midway point in the beamline, the beam has yet to be significantly collimated and therefore is largely divergent: this is reflected in the differences between the penumbra regions. 

\begin{figure}[htb]
    \centering
    \begin{subfigure}[b]{\textwidth}
    \centering
    \includegraphics[width=0.72\textwidth]{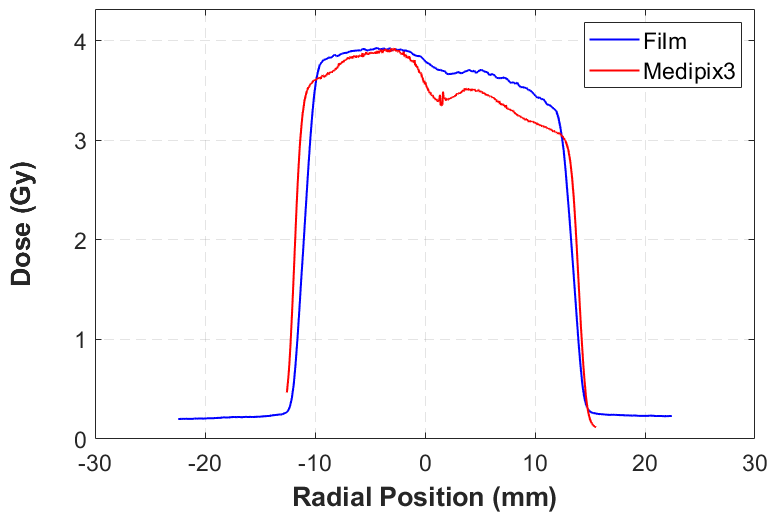}
    \caption{}
    \end{subfigure}
    \vfil
    \begin{subfigure}[b]{\textwidth}
    \centering
    \includegraphics[width=0.72\textwidth]{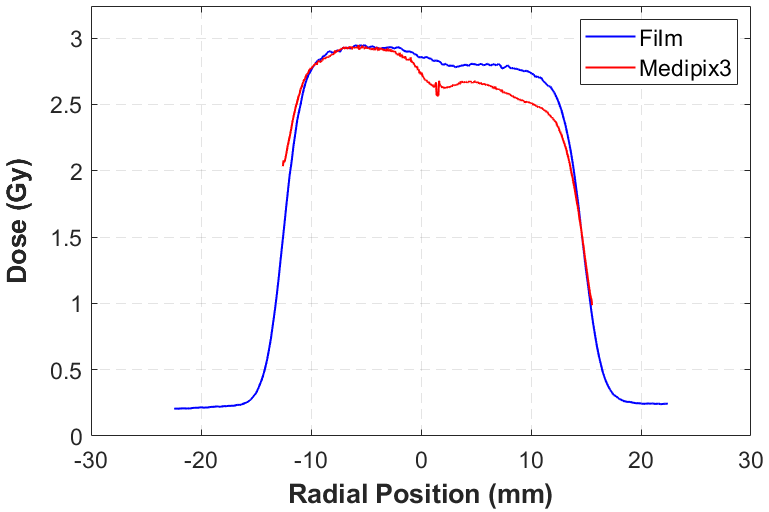}
    \caption{}
    \end{subfigure}
  \caption{Beam profiles obtained with film and Medipix3 at (a) 9.5 cm and (b) 30 cm after the treatment nozzle.}
    \label{fig:profilesNozzle}
\end{figure}

Measurement runs 6 and 7 with the system placed at 9.5 cm and 30 cm after the treatment nozzle are shown in figure~\ref{fig:profilesNozzle}. A slightly wider profile was attained at 30 cm as the beam spreads out further downstream. Better agreement between the distributions was achieved for these cases, particularly at the lateral penumbra. A tilt in the same direction is also observed here, although the resulting profiles are less imbalanced. A flat, rectangular distribution is expected here but evidently, even a slight angular deviation results in an asymmetrical profile which is easily detectable by both methods. 

There are several considerations for these irregularities which are related to the image analysis, beam stability and quality on the day, as well as unknowns within the delivery system. It was found that the selection of the ROI in ImageJ was important and can affect the shape and resolution of the profile. As the beam distribution at the first position differs considerably to measurements after the nozzle, a rectangle ROI was chosen. This was specified such that there was total horizontal beam coverage and sufficient height to generate a smooth profile: the profile smears out if the ROI is too large. These ROI settings were kept the same for all images. 

For all of the film profiles (figures~ \ref{fig:profilesIntegZone}--\ref{fig:profilesNozzle}) the tail regions do not extend to 0 Gy, this is due to the chosen fit for the calibration curves. We did not correct for this as primarily the central regions and highest points of dose were considered for comparison with the Medipix3 detector. Furthermore, at these lower levels of dose ($<$5 Gy), there is a small margin of difference given the exponential fit derived from the calibration curve (figure \ref{fig:calibrationcurve}). 

Nonetheless, the film provided an intermediary means to correlate dose at all these positions to the ionisation chambers: these are indicative doses rather than precise measurements due to operational uncertainties described in section \ref{FilmAnalysis}. The limitations surrounding the analysis and conversion process are indicated by the error bars in the calibration curves. A way to mitigate these would be to proceed with a more robust process, by strictly monitoring the film protocol to maintain uniformity with the film irradiation, development, scanning and analysis; or to attain profiles using well established or commercial methods. For example, using commercial software and hardware to determine the beam profiles from the irradiated film or by direct measurements with standard profile monitors or ionisation chambers. The possibility of determining the dose from counts is further described in section \ref{section:detector-response}.

In addition to any misalignment of the system, it is also possible that small beam offsets at the start of the treatment line or aberrations with the shape or positioning of the beam stopper could impact the resulting beam distributions. This is less of an issue for the actual delivery of treatment as the beam is well conditioned to operate within the stable clinical currents with acceptable parameters, maintained by the QA diagnostics. Furthermore, due to the design of the delivery system and drift distance ($\sim$1.5 m) to the nozzle, further fluctuations are moderated by operational changes with the transport line optics. For these measurements, several issues were encountered as the beam currents requested were lower than the clinical range and the beam itself was not regulated as required for treatment.

\subsection{Detector response}
\label{section:detector-response}
The comparative profiles demonstrate similar capabilities to provide spatial measurements, particularly at the penumbra regions. However, these measurements could be considered to be a `worst case' scenario for the detector due to the relatively thick silicon sensor and the charge sharing effect contributing to a guaranteed increase in count rate. Both effects increase the difference between expected and measured count rate.

An estimation of the magnitude of effective charge sharing for single protons was made by using the ImageJ Shape Filter plugin which uses the IJBlob library \cite{Wagner2013}. This identifies clusters and returns many cluster properties such as area, aspect ratio, circularity etc. At the beginning of the measurements (between runs 2--3) using 967 identified clusters, the mean cluster area was 7.6 pixels, the mode area was 4 pixels with 187 counts and the minimum area was 3 pixels. The central part of the image was excluded for this first analysis because the beam was approximately centred on the detector which would have skewed the analysis. By the end of the measurement (run 11) using 1874 identified clusters, the mean cluster area increased to 9.3 pixels, the mode area decreased to 3 pixels with 348 counts and the minimum area stayed at 3 pixels.

This is evidence that the measured count rate by the Medipix3 could be corrected for by dividing the count rate by a factor between 3--9.3. In order to arrive at a more precise and reliable correction factor, dedicated measurements with very low flux of protons with precisely known energy would be necessary.

The minor detector artefacts observed in the central and cross pixels are due to a larger than average effective pixel size for those pixels, approximately a factor of 3 times larger. This varies depending on the exact manufacturing process which improves over time by shrinking this gap. In order to compensate for the larger pixel size with x-rays, a first  order correction is to divide the raw counts by 2.8. However, the response of cross and central pixels cannot be very well corrected for by a linear scaling since there are non-linear effects including the count rate linearity. Therefore, only a basic linear scaling was applied to approximately match the counts of neighbouring pixels.

\begin{figure}[htb]
    \centering
    \includegraphics*[width=0.6\textwidth]{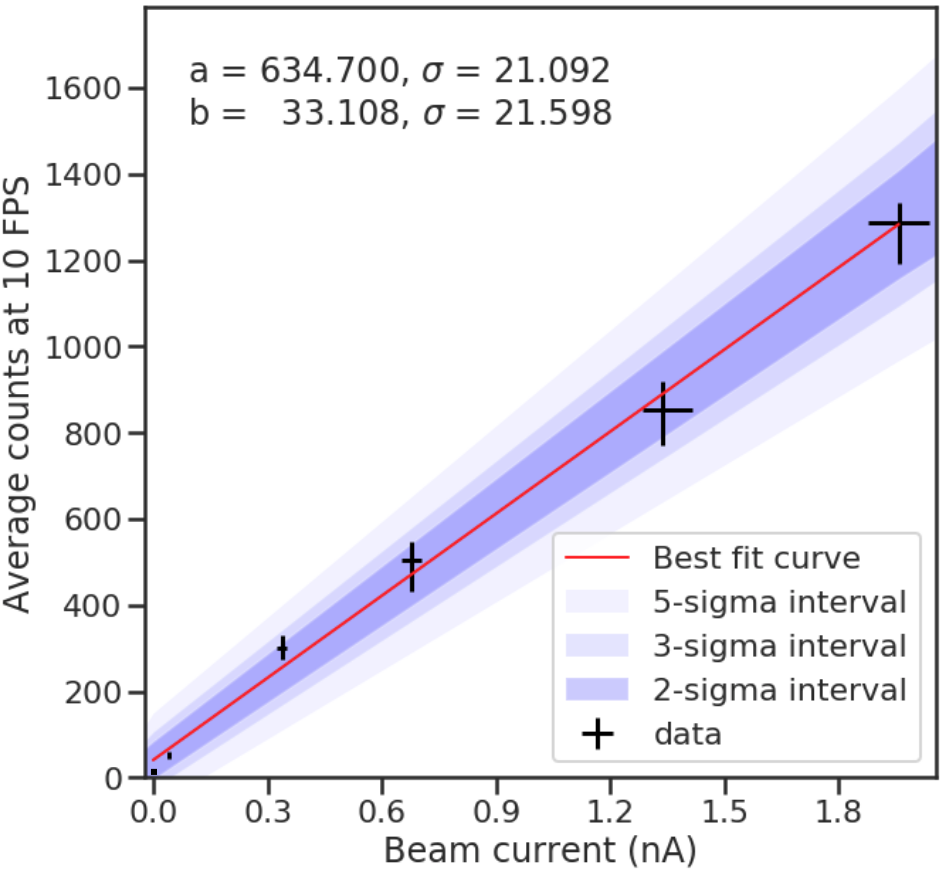}
    \caption{Count rate linearity over all active pixels recorded at 10 FPS (frames per second) for 6 beam currents between $0.012_{-9\%}^{+13\%}$ and $1.97_{-7\%}^{+4\%}$ nA.}
    \label{fig:countrate}
\end{figure}

Figure~\ref{fig:countrate} shows that the detector has a linear response across the entire tested range of beam currents from 0.012 to 1.97 nA. There is relatively large uncertainty of the average count rate due to the electrometer beam current measurements, dominated by the infrequent, manual readings. This is included in the residuals of the data and the variation in beam current which are both approximately in the order of $10\%$. Further measurements with improved beam current control would significantly reduce the count rate uncertainty and thus provide a reliable demonstration of count rate linearity with the Medipix3 detector. Nevertheless, the ultra low beam currents from runs 2 and 3 would not be possible to measure with other commonly used instruments with the precision of single protons; this detector therefore enables semi-destructive beam current measurements from single events to $10^{10}$ protons per second with the possible temporal resolution down to 0.5 ms.

The Medipix3 detector appears to have sufficient count rate linearity and sensitivity for beam characterisation and quality assurance measurements. Once cross-checked with another detector for accurate verification of the beam current, the impact of the `charge sharing' effect on absolute dosimetry could be quantified accurately. A first order correction would be to simply divide the counts by the average cluster size, for these measurements this value would be 3--9.3 as mentioned in section \ref{section:detector-response}. It is also anticipated that the Medipix3 chip configuration could be optimised significantly for high flux, clinical energy protons from the default x-ray (4--30 keV) settings. For future tests, if the count rates are low enough, it would be relevant to use the `charge summing mode' on Medipix3 for furthering detector characterisation and possible mitigation of charge sharing. This has been shown to work as intended with relatively low energy x-rays \cite{Gimenez_2011}, however the energy deposition from these 60 MeV protons is significantly higher, and so needs verification. Further work in this area could determine the energy deposited in the sensor and obtain definitive conversion factors to determine absorbed dose directly from the induced charge. This would also enable direct comparison with the dose distributions obtained by the film. 

\subsection{Detector activation}
\label{DetectorActivation}

An overview of five background measurements during and between measurement runs is shown in table \ref{tab:detectorActivation}. These quantify how much the detector was activated over time, including the standard deviation, start and end count rates. The main count rate with the beam off is shown to increase over the course of the measurement day from 0 counts per second before the beam irradiation commenced to 1.3 counts per second at the end. The final measurement (run 15) was taken during the a proton beam interruption lasting 94 ms and so is different to the other measurements. The average count rate of 1.3 counts per second for this run is therefore indicative of the instantaneous detector activation between full beam intensity at 2.1 nA. This is relevant as a number of excited states have short half lives in the millisecond to second range which would not be evident in the other 4 measurements shown in this table.

\begin{table}[htb]
\centering
\caption{Overview measurement parameters of detector activation in counts per second with acquisition times and run ranges or numbers. The rows showing ranges of run numbers indicate the measurement was between run numbers. }
\begin{tabular}{|c|c|c|c|c|c|}
\hline
\shortstack{Run (\#)\\$\;$} & \shortstack{Time (s)\\$\;$} & \shortstack{Mean (counts/s)\\$\;$} & \shortstack{\\Standard deviation \\(counts/s)} & \shortstack{Start (counts/s)\\$\;$} & \shortstack{End (counts/s)\\$\;$} \\ \hline
2--3 & 50    & $7.2 \times 10^{-4}$ & $1.3 \times 10^{-3}$ & $9.5 \times 10^{-4}$ & $6.7 \times 10^{-4}$ \\
5--6 & 20    & $4.4 \times 10^{-2}$ & $2.8 \times 10^{-2}$ & $4.7 \times 10^{-2}$ & $4.0 \times 10^{-2}$ \\
8--9 & 10    & $6.4 \times 10^{-2}$ & $3.8 \times 10^{-2}$ & $6.7 \times 10^{-2}$ & $6.1 \times 10^{-2}$ \\
11    & 8.3   & $9.8 \times 10^{-2}$ & $6.4 \times 10^{-2}$ & $9.9 \times 10^{-2}$ & $9.7 \times 10^{-2}$ \\
15    & 0.094 & 1.3                  & 0.63                 & 1.5                  & 1.2                  \\ \hline
\end{tabular}
\label{tab:detectorActivation}
\end{table}

\subsection{Temporal analysis of the beam}
\label{TemporalAnalysis}
A linear response was observed across the current range and the beam performance was reliable at higher clinical dose rates. Investigating the stability of the beam flux is a novelty in itself as it is not possible to measure the beam profile in real-time with millisecond resolution with other methods such as EBT3 film, beam current monitors or wire scanners. The beam current monitors could be cross-checked with these data if it was possible to log their outputs at high speed, the control system did not allow such control. These measurements verified that the beam profile was stable over time with a variable flux during a measurement. However this does not significantly affect the clinical treatment as CCC uses passive modulation with patient specific collimators for dose distribution control rather than a scanning pencil beam.

\begin{figure}[htb]
    \centering
    \includegraphics[width=0.78\columnwidth]{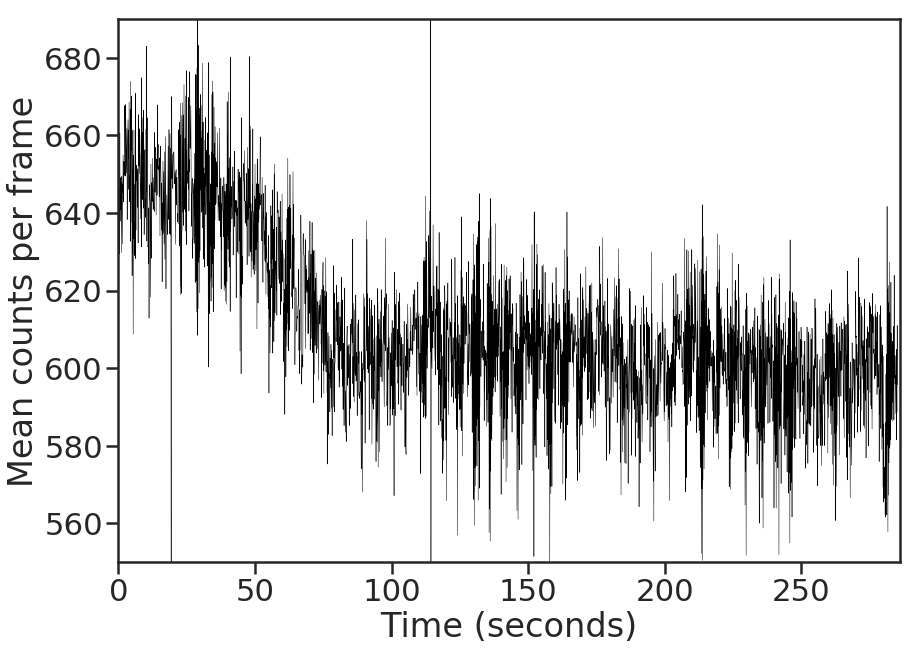}
    \caption{Temporal stability of the beam over all pixels over nearly 5 minutes, recorded at 100 FPS, 26.3 Gy/minute in the integration zone.}
    \label{fig:temporal_stability_of_beam}
\end{figure}

Figure~\ref{fig:temporal_stability_of_beam} shows the mean counts integrated over all pixels over the typical measurement times for this work (a few minutes), at 100 FPS. The mean counts per frame are directly proportional to the instantaneous dose for a 10 ms period. The variations in the mean counts per frame show that the measured counts are not totally stable over time, the source of the instability could be due to the detector or the accelerator. The mean counts per frame varied over a measurement from the start to end by approximately 50 counts out of between 600 and 660 counts, which is ~8\%. However, recurring jitter were seen in the data where variations in the count rate over time appeared as small amplitude waves at low (few ms) and higher (of the order of hundreds of ms) frequencies. The mean counts per frame were qualitatively observed to oscillate at at least one higher frequency during the measurements. These peculiarities were quantified with a frequency decomposition analysis.

\begin{figure}[htb]
    \centering
    \includegraphics[width=0.7\columnwidth]{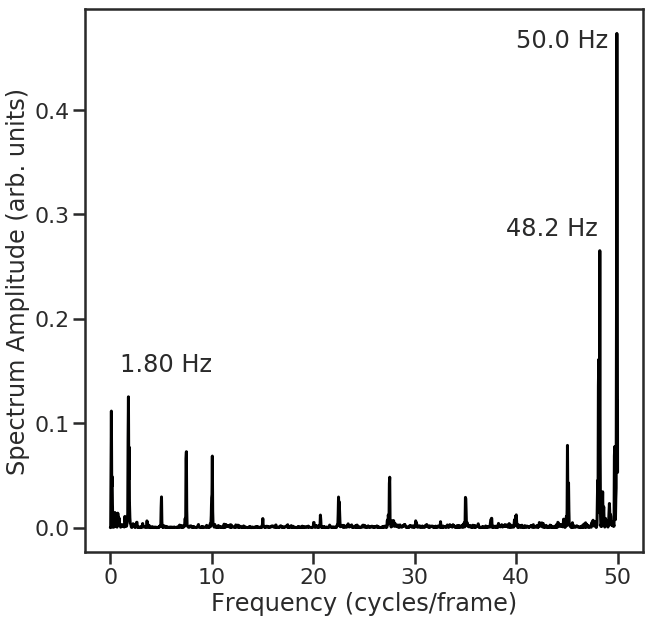}
    \caption{Frequency components of the beam intensity as recorded by all pixels over nearly 5 minutes recorded at 100 FPS from figure~\ref{fig:temporal_stability_of_beam}. The 50.0 and 48.2 Hz components are the strongest followed by a 1.80 Hz component.}
    \label{fig:frequency_components_temporal_stability_of_beam}
\end{figure}

In figure~\ref{fig:frequency_components_temporal_stability_of_beam}, the outliers are indicated by the three labelled frequency peaks. The 50.0 Hz component is assumed to be related to the UK AC mains electricity as it exactly matches the frequency. The 48.2 Hz component was not experienced during previous detector tests with x-rays or seen with any other measurements at the beamline. It is unclear if it is associated with the accelerator or detector: further measurements are necessary to identify a candidate. It is hypothesised that the 1.80 Hz component is related to the ion source. As mentioned in section \ref{FilmAnalysis}, the change to a proton therapy service required the nominal arc current (of the order of mA) of the ion source to be limited such that the beam currents produced were practical for treatment (nA range). Only the 50 Hz component was observed in later measurements performed at MedAustron \cite{Bal2021} which is a significantly different facility and operates a synchrotron. As such, it is presumed that these other frequency peaks are specifically associated with the CCC cyclotron or may be due to interactions with beamline elements. 

The significance of these rapid oscillations may be meaningful as they demonstrate something otherwise unforeseen which may have an impact on beam performance. For example, in passive delivery at CCC and the beam is modulated with a rotating wheel to generate a spread out Bragg peak, the graduations determine the longitudinal distribution of dose. Variations in the beam current could cause asynchronicity with the rotational frequency of the wheel and aberrations with the dose delivered for each step. 

If the oscillations are present in regular operation, these measurements may reveal an underlying aspect of the cyclotron. An irregularity in the structure or accelerating process can influence the beam dynamics (orbit, stability, focusing effects from the magnetic field etc.) of the circulating protons. This may in turn impact the extracted beam quality such as the energy spread, emittance and dispersion. The jitter is presumed to be caused by a combination of different factors including the design, adjustments with the ion source and aged components. It is unknown if or how the beam is adjusted upon extraction. If there is a collimator at the exit then some properties of the beam would be regulated, such as the beam distribution in the transverse plane. 

This analysis poses interesting avenues to study the beam structure and dynamics of accelerated CPT beams with the Medipix3 detector, for routine measurements or commissioning purposes. The timing capabilities would enable the possibility of examining the beam intensities, quality and stability of both continuous and pulsed beams. This would be also applicable for active delivery schemes such as pencil beam scanning. 

Subsequent studies could investigate track projection over multiple pixels and the charge sharing as a function of beam angle relative to the detector. Efforts could be made to minimise the material in the beam path so that the detector becomes less activated over time; for this work, the large aluminium heatsink would have been the majority of the activated material and could be significantly reduced. Solutions could include a high RPM fan blowing air or dry ice over the detector or transferring the heat generated via vapor chambers, heat pipes or pyrolytic graphite to a heatsink just outside of the main beam path.

The detector settings could be significantly improved upon by tuning for high energy protons instead of the default, low energy x-rays. There are various software configurable DACs (digital-to-analogue converters) to control parameters which control front-end signal rising and falling times, signal baseline levels, thresholds among many others some of which are specifically designed with high count rates in mind like `Pole Zero Cancellation' ($R_{PZ}$). In the same direction, ideally the detector would have a very thin sensor in order to induce much less charge sharing and still measure a very high amplitude signal. This would also reduce the detector activation as discussed in section \ref{DetectorActivation}.

\section{Conclusions and Outlook}
Significant innovations in technology have resulted in greater accessibility and benefits of CPT worldwide. However, rapid developments in recent years have highlighted the need for improved diagnostic systems and performance capabilities, to fulfil the requirements of emerging, advanced delivery techniques. There is ongoing interest in the application of silicon detectors and in this work, we examined the potential of the Medipix3 hybrid pixel detector for CPT. 

Experimental measurements were performed at the CCC 60 MeV clinical proton therapy beamline and compared with standard film dosimetry methods. Simultaneous irradiation of the film and detector placed at multiple locations in the delivery system allowed a direct comparison of the transverse beam distributions. There was general agreement between both methods, particularly at the lateral edges of the beam although several uncertainties resulted in experimental irregularities, also influenced by beam quality on the day. 

Medipix3 has the capacity to measure the beam current by the detection of individual protons with millisecond scale temporal resolution and almost instantaneous readout time. This enabled the possibility to resolve otherwise unknown information about the CCC beam and accelerator, undetectable with typical instruments. 

To facilitate its progression toward clinical implementation, further testing is recommended to characterise the cluster properties, signal uniformity, sensitivity across the detector, activation levels, dose rate thresholds, energy dependence, stability, spatial resolution and dosimetric calibration factors. These should be performed at different facilities, across the full treatment energy range and with different particle types. 
 
We present the first set of tests using the Medipix3 detector technology in a clinical proton beam environment. This work explores the promising capabilities and versatility of Medipix3 for CPT, suggesting its possibility as a fast and efficient, future tool for routine dosimetry, commissioning and beam monitoring. 

\acknowledgments
This project was supported by funding from the European Union FP7 grant agreement 215080, H2020 research and innovation programme under the Marie Sk\l{}odowska-Curie grant agreement No 675265, OMA – Optimization of Medical Accelerators and the Cockcroft Institute core grant STGA00076-01. This work is part of the research programme of the Foundation for Fundamental Research on Matter (FOM), which is part of the Netherlands Organisation for Scientific Research (NWO). It was carried out at the National Institute for Subatomic Physics (Nikhef) in Amsterdam, the Netherlands.

\bibliographystyle{JHEP}
\bibliography{references}

\end{document}